\documentclass[prl,twocolumn,showpacs]{revtex4}

\usepackage{graphicx}
\usepackage{dcolumn}
\usepackage{bm}

\begin{document}
\title{Diffusion and Relaxation Controlled by Tempered $\alpha$-stable Processes}

\author{Aleksander Stanislavsky}
\affiliation{Institute of Radio Astronomy, 4 Chervonopraporna St., 61002 Kharkov,
Ukraine} \email{alexstan@ri.kharkov.ua}
\author{Karina Weron}%
\affiliation{Institute of Physics,  Wroc{\l}aw University of Technology, Wyb.
Wyspia$\acute{n}$kiego 27, 50-370 Wroc{\l}aw, Poland}%
\email{karina.weron@pwr.wroc.pl}
\author{Aleksander Weron}
\affiliation{Hugo Steinhaus Center, Institute of Mathematics and Computer Science,
Wroc{\l}aw University of Technology, Wyb. Wyspia\'{n}skiego 27, 50-370 Wroc{\l}aw,
Poland} \email{aleksander.weron@pwr.wroc.pl}
\date{\today}

\begin{abstract}
We derive general properties of anomalous diffusion and nonexponential relaxation from
the theory of tempered $\alpha$-stable processes. Its most important application is to
overcome the infinite-moment difficulty for the $\alpha$-stable random operational time
$\tau$. The tempering results in the existence of all moments of $\tau$. The
subordination by the inverse tempered $\alpha$-stable process provides diffusion
(relaxation) that occupies an intermediate place between subdiffusion (Cole-Cole law) and
normal diffusion (exponential law). Here we obtain explicitly the Fokker-Planck equation,
the mean square displacement and the relaxation function. This model includes
subdiffusion as a particular case.
\end{abstract}

\pacs{05.40.Fb, 02.50.Ey, 77.22.Gm}
\maketitle

{\it Introduction.}--- Many studies have been reported on the phenomenon of subdiffusion
which is typically observed when, due to dominating influence of traps (see
\cite{MK,zasl02,SK} and reference therein), the waiting times of random walks become
$\alpha$-stable, $\alpha < 1,$ with an infinite mean. However, this picture is only an
idealization of the physical world. In reality the time of trap life can be restricted.
It should be taken into account that the traps can be located in some spatial regions
from which a walker may easily escape. Consequently, in a more general representation,
the random walks start as subdiffusion, but their characteristics become very similar to
those of normal diffusion at large time scales. One of such clear examples is a random
motion of bright points (BP) associated with magnetic fields at the solar photosphere.
The BP transport in the intergranular lanes with times less than 20 minutes has a
subdiffusive character, but the analysis of the BP motion supports the normal diffusion
behavior for larger times. The experimental result is reported in \cite{Cad}. The present
paper is just devoted to such a problem. For this purpose we are going to apply the
tempered $\alpha$-stable processes for the description of diffusion and relaxation. In
comparison with the purely $\alpha$-stable process such a process has finite moments, but
it saves some important rudiments of the stable process too \cite{Ros07,PirSW,Terdik}.
Therefore, if its inverse process is taken as a subordinator, it provides then a
diffusive picture occupying an intermediate place between subdiffusion and normal
diffusion. 

{\it Tempered $\alpha$-stable process and its inverse.}--- The model of subdiffusion is
based on a flexible Montroll-Weiss idea on continuous time random walks (CTRW)
\cite{Montroll}. Briefly, the representation of anomalous diffusion by means of the CTRW
methodology is the following (for more details see, for example, \cite{S0,MW1}). Consider
a sequence ${T_i}$, $i=1,2,\dots$ of non-negative, independent, identically distributed
(iid) random variables which represent waiting-time intervals  between subsequent jumps
of a walker. The random time interval of $n$ jumps in space equals $T(n)=\sum_{i=1}^n
T_i$ with $T(0)$ = 0.
The random number $N_t$ of jumps, performed by the walker up to time $t>0$, is determined
by the largest index $n$ for which the sum of $n$ interjump time intervals does not
exceed the observation time $t$, namely $N_t=\max\{n:T(n)\leq t\}$.
The position of the walker after $N_t$ jumps becomes then
\begin{equation}
R(N_t)=\sum_{i=1}^{N_t}R_i,\label{eqa11a}
\end{equation}
where $R_i$ are iid variables giving both the length and the direction of the {\it i}-th
jump. The process (\ref{eqa11a}) is just known as the
CTRW. 

If the time intervals $T_i$ belong to the domain of attraction of a completely asymmetric
$\alpha$-stable distribution with the index $0<\alpha<1$, the generalization of the
central
limit theorem yields the continuous limit 
$a^{-1/\alpha}T([a\tau])\stackrel{d}{\rightarrow}U_\tau$ as $a\to\infty$,
where $U_\tau$ is a strictly increasing $\alpha$-stable L$\acute{\rm e}$vy process, $a>0$
parameter, $[x]$ denotes the integer part of $x$ and ``$\stackrel{d}{\rightarrow}$''
means ``tends in distribution''. Similarly, let the jumps $R_i$ belong to the domain of
attraction of a $\gamma$-stable distribution $S_{\gamma,\beta}(x), 0<\gamma\leq 2,
|\beta|\leq 1$ so that $a^{-1/\gamma}R([a\tau])\stackrel{d}{\rightarrow}X_\tau$ as
$a\to\infty$,
where $X_\tau$ is a $\gamma$-stable L$\acute{\rm e}$vy process known as the parent
process. If $\gamma=2$, the parent process is the classical Brownian motion. Both, the
process $U_\tau$ and the process $X_\tau$ are indexed by random operational (internal)
time $\tau$. In order to find a particle position at the observable time $t$, we have to
introduce the notion of the inverse-time $\alpha$-stable subordinator $S_t$ relating the
internal and the observable times
\begin{equation}
a^{-\alpha}N_{at}\stackrel{d}{\rightarrow}S_t=\inf\{\tau:U_\tau >t\}\qquad \label{eqa12}
\end{equation}
as $a\to\infty$. Then, as $a\to\infty$, the continuous limit of the CTRW process
(\ref{eqa11a}) obtains the following form
\begin{equation}
a^{-\alpha/\gamma}R(N_{at})\approx(a^\alpha)^{-1/\gamma}R([a^\alpha
S_t])\stackrel{d}{\rightarrow}X(S_t) \label{eqa13},
\end{equation}
known as the anomalous diffusion process \cite{Meer}, directed by the inverse
$\alpha$-stable subordinator $S_t$. It should be pointed out that the process $U_\tau$
does not have any finite $p$-moments for $p\leq \alpha$. Therefore, the subdiffusion is
characterized by a power mean square displacement in time \cite{S0,MW1,S2}.

However, there are physical phenomena, for example, the random motion of BPs in
intergranular lanes on the Sun, where it would be desirable to get also a model that
overcomes the infinite-moment difficulty while preserving the subdiffusive behavior for
short times \cite{StWer07}. The remedy was first proposed in the physical literature by
Mantegna and Stanley \cite{ManStan}. Their idea of truncated L\'evy flights served as a
model for random phenomena which exhibit at small scales properties similar to those of
L\'evy flights, but have distributions which at large scales have cutoffs and thus have
finite moments of any order. Koponen \cite{Koponen}, building on Mantegna and Stanley's
ideas, defined the smoothly truncated L\'evy flights which stressed the advantage of a
nice analytic form. Independently, the same family of distributions was described earlier
by Hougaard \cite{Houg} in the context of a biological application. However, different
methods for the truncation were suggested  also in the economic and statistical sciences
\cite{Boyarchenko, Carr, Kim}, but until the Rosi\'nski's paper \cite{Ros07} there was a
lack of invariance under linear transformations for the distributions introduced, a
significant property that the $\alpha$-stable laws possess. He succeeded in finding the
appropriate class of tempered stable distributions and processes \cite{Terdik, Kim}.

In what follows we discuss properties of a diffusion process which is related to an
inverse tempered $\alpha$-stable subordinator. The Laplace image for the probability
density function  (p.d.f.) of a tempered non-negative $\alpha$-stable variable is
\begin{equation}
\tilde{f}(u)=\exp\left(\delta^\alpha-(u+\delta)^\alpha\right) \,\label{eq5},
\end{equation}
where $\delta$ is a positive constant and $0 < \alpha < 1$  \cite{PirSW}. If $\delta$
equals zero, the tempered $\alpha$-stable p.d.f. becomes simply $\alpha$-stable.
Eq.(\ref{eq5}) describes probabilistic properties of the tempered $\alpha$-stable L\'evy
process $U(\tau)$ which generalizes the above mentioned process $U_\tau$.

Next, we will find its inverse process $S(t)$ as in (\ref{eqa12}), where $U(\tau)$
substitutes $U_\tau$. If $f(t,\tau)$ is the p.d.f. of $U(\tau)$, then the p.d.f.
$g(\tau,t)$ of its inverse $S(t)$ can be represented as
\begin{displaymath}
g(\tau,t)=-\frac{\partial}{\partial\tau}\int_{-\infty}^tf(t',\tau)\,dt'.
\end{displaymath}
Taking the Laplace transform of $g(\tau,t)$ with respect to $t$, we get
\begin{equation}
\tilde{g}(\tau,u)=\frac{(u+\delta)^\alpha-\delta^\alpha}{u}
\,e^{-\tau[(u+\delta)^\alpha-\delta^\alpha]} \,.\label{eq6}
\end{equation}
When $u\gg 1$ ($t\ll 1$) or $\delta\to 0$, Eq. (\ref{eq6}) tends to
\begin{equation}
\tilde{g}(\tau,u)=u^{\alpha-1}\,e^{-\tau u^\alpha}\,,\label{eq6a}
\end{equation}
which is the Laplace image of an inverse $\alpha$-stable p.d.f. typical for subdiffusion.
If $u\ll 1$ ($t\gg 1$) or $\alpha\to 1$, then Eq. (\ref{eq6}) becomes the Laplace image
of the Dirac delta-function. It follows from Eq. (\ref{eq6a}) that the p.d.f. of the
inverse $\alpha$-stable process is
\begin{displaymath}
g(\tau,t)=\frac{1}{2\pi i}\int_{Br}e^{ut-\tau
u^\alpha}\,u^{\alpha-1}\,du=t^{-\alpha}F_\alpha(\tau/t^\alpha)\,,
\end{displaymath}
where $Br$ denotes the Bromwich path, and the function $F_\alpha(z)$
is a specific case of the Wright function \cite{Erd,S0}.


{\it Subordination by an inverse tempered $\alpha$-stable process.}--- Let the parent
process $X(\tau)$ have the p.d.f. $h(x,\tau)$. Then the p.d.f. of the subordinated
process $X[S(t)]$ obeys the integral relationship between the pro-bability densities of
the parent and directing processes, $X(\tau)$ and $S(t)$, respectively,
\begin{equation}
p(x,t)= \int^\infty_0h(x,\tau )\,g(\tau,t)\,d\tau. \label{eq8}
\end{equation}
In the Laplace space the probability density $p(x,t)$ has the most simple form. Taking
into account Eq. (\ref{eq6}), the Laplace transform of Eq. (\ref{eq8}) with respect to
$t$ gives
\begin{equation}
\tilde{p}(x,u)= \frac{(u+\delta)^\alpha-\delta^\alpha}{u}\,
\tilde{h}(x,(u+\delta)^\alpha-\delta^\alpha). \label{eq9}
\end{equation}
For $\delta=0$ the latter expression becomes $u^{\alpha-1}\tilde{h}(x,u^\alpha)$.

\begin{figure}
\includegraphics[width=8.6 cm]{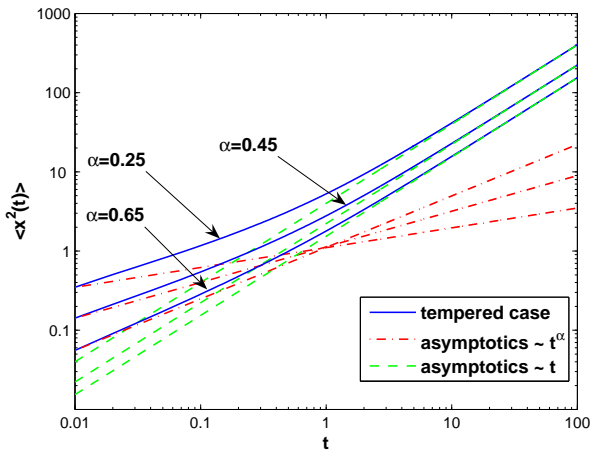}
\caption{\label{fig:diff}(Color online) Mean square displacement of anomalous diffusion
subordinated by an inverse tempered $\alpha$-stable process.}
\end{figure}

It is not difficult to calculate the moments of the process $X[S(t)]$ if the moments of
the process $X(\tau)$ are known. For example, for the Gaussian process ($\gamma = 2$) the
second moment is $\left\langle X^2(\tau)\right\rangle=D\tau$, where $D$ is a diffusive
constant. Then the mean square displacement of $X[S(t)]$ can be written as
\begin{displaymath}
\left\langle X^2[S(t)]\right\rangle=\int^\infty_0\left\langle
X^2(\tau)\right\rangle\,g(\tau,t)\,d\tau\,.
\end{displaymath}
The Laplace image $\tilde{\left\langle X_u^2\right\rangle}$ of $\left\langle
X^2[S(t)]\right\rangle$ has the form
\begin{equation}
\tilde{\left\langle
X_u^2\right\rangle}=\frac{D}{u[(u+\delta)^\alpha-\delta^\alpha]}\,.\label{eq10}
\end{equation}
Consequently, the inverse Laplace transform of Eq. (\ref{eq10}) reads
\begin{equation}
\frac{1}{u[(u+\delta)^\alpha-\delta^\alpha]}\stackrel{{\cal
L}^{-1}}{\rightarrow}\int_0^te^{-\delta
y}\,y^{\alpha-1}\,E_{\alpha,\,\alpha}(\delta^\alpha y^\alpha)\,dy\,,\label{eq11}
\end{equation}
where
\begin{displaymath}
E_{\alpha,\,\beta}(z)=\sum_{k=0}^\infty\frac{z^k}{ \Gamma(\alpha k+\beta)}, \quad
\alpha>0,\quad \beta>0
\end{displaymath}
is the Mittag-Leffler function \cite{Erd}. The function (\ref{eq11}) gives rise to
interesting asymptotic properties of the mean square displacement $\left\langle
X^2[S(t)]\right\rangle$. For $t\ll 1$ this displacement behaves as
$Dt^\alpha/\Gamma(\alpha+1)$, but for $t\gg 1$ it increases linearly in time $Dt/\alpha$
(see Fig.~\ref{fig:diff}). Thus the anomalous diffusion, governed by the inverse tempered
$\alpha$-stable subordinator, occupies an intermediate place between sub-diffusion and
the normal diffusion. For short times it behaves as subdiffusion whereas for the long
times it resembles the properties of the normal diffusion. Let us call the diffusion
subordinated by the inverse tempered $\alpha$-stable process as a ``tempered
subdiffusion''. As is well known \cite{bbm05},
the inverse $\alpha$-stable process accounts for the amount of time, when a walker does
not participate in a motion.
By analogy, we may conclude that the process $S(t)$ for the tempered subdiffusion
represents a case, when a walker does not participate in a motion only for restricted
intervals of time. At large time scales the walker begins to move randomly without any
stopping as if $\alpha=1$.

{\it Equation of tempered subdiffusion.}--- Let $\hat L(x)$ be a time-independent
Fokker-Planck operator (FPE), whose exact form is not important here. Let the ordinary
FPE $\partial h(x,\tau)/\partial\tau=\hat L(x)\,h(x,\tau)$ describe the evolution of a
particle subject to the operation time $\tau$. Acting by the operator $\hat L(x)$ on the
image $\tilde{p}(x,u)$ from Eq. (\ref{eq9}), we find
\begin{eqnarray}
\hat L(x)\,\tilde{p}(u,x)&=&[(u+\delta)^\alpha-\delta^\alpha]\,\tilde{p}(x,u
)-\nonumber\\ &-&q(x)\,\frac{[(u+\delta)^\alpha-\delta^\alpha]}{u}\,,\label{eq12}
\end{eqnarray}
where $q(x)$ is an initial condition. When $\delta=0$, the inverse Laplace transform of
the latter expression gives a fractional representation of the FPE \cite{MK, S0}
\begin{eqnarray}
p(x,t)&=&q(x)+\nonumber\\
&+&\frac{1}{\Gamma(\alpha)}\int_0^td\tau\,(t-\tau)^{\alpha-1}\,\hat
L(x)\,p(x,\tau)\,.\label{eq13}
\end{eqnarray}
In the case of the tempered subdiffusion the kernel in the integral representation of the
FPE will be more complex, containing as a special case the kernel of Eq.(\ref{eq13}) for
$\delta\to 0$. Using the formal integral representation of the FPE
\begin{equation}
p(x,t)=q(x)+\int_0^td\tau\,M(t-\tau)\,\hat L(x)\,p(x,\tau)\,.\label{eq14}
\end{equation}
and taking the inverse Laplace transform of Eq.(\ref{eq12}), we obtain the explicit form
of the kernel $M(t)$, namely
\begin{equation}
M(t)=e^{-\delta t}\,t^{\alpha-1}\,E_{\alpha,\,\alpha}(\delta^\alpha
t^\alpha)\,.\label{eq15}
\end{equation}
For $t\ll 1$ (or $\delta\to 0$) this function takes the power form
$t^\alpha/\Gamma(\alpha)$ as the kernel in Eq. (\ref{eq13}). However, for  $t\gg 1$ (or
$\alpha\to 1$) $M(t)$ becomes constant and, as a result, Eq. (\ref{eq14}) transforms into
the integral form of the ordinary FPE.

{\it Tempered relaxation.}---The commonly accepted theoretical approaches to model
relaxation phenomena assume \cite {MK} decay of an excitation undergoing diffusion in the
system under consideration. In this framework, the relaxation function $\phi(t)$
describes the temporal decay of a given mode $k$ and can be expressed \cite {MW1} through
the Fourier transform of the diffusion process $X[S(t)]$
\begin{equation}
\phi(t)=\left\langle e^{-kX[S(t)]}\right\rangle\,.\label{eq16}
\end{equation}
Here $k>0$ has the physical meaning of a wave number (the Fourier image of spatial
coordinates). Starting with Eq. (\ref{eq6}), we can write the Laplace image of Eq.
(\ref{eq16}) as
\begin{equation}
\tilde{\phi}(u)=\frac{[(u+\delta)^\alpha-\delta^\alpha]}{u[\Phi(k)+(u+\delta)^\alpha-\delta^\alpha]}\,,\label{eq17}
\end{equation}
where $\Phi(k)$ is the logarithm of the characteristic function of the process $X(\tau)$.

\begin{figure}
\includegraphics[width=8.6 cm]{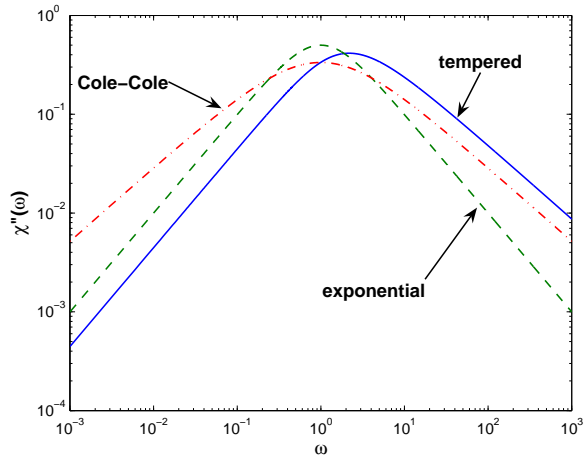}
\caption{\label{fig:relax}(Color online) Absorption (imaginary) term of the
frequency-domain relaxation function $\chi(\omega)=\chi'(\omega)-i\chi''(\omega)$ for
$\alpha$ = 0.75\,.}
\end{figure}

\begin{figure}
\includegraphics[width=8.6 cm]{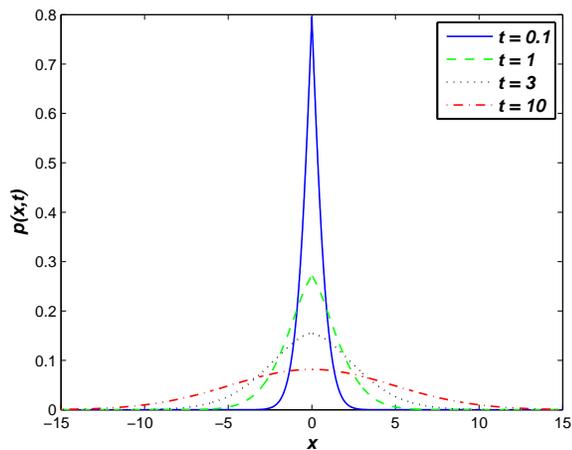}
\caption{\label{fig:prop}(Color online) Propagator $p(x,t)$ for the tempered diffusion
with $\alpha$ = 2/3 and $\delta$ = 0.5, drawn for consecutive dimensionless instances of
time $t$ = 0.1, 1, 3, 10. The cusp shape of the p.d.f. disappears.}
\end{figure}

To expose the characteristic properties of the ``tempered relaxation'' we use the
frequency-domain description \cite{J2, Jurlewicz} of the relaxation phenonenon
\begin{equation}
\chi(\omega)=\int^\infty_0e^{-i\omega
t}\,\left(-\frac{d\phi(t)}{dt}\right)\,dt\,.\label{eq18}
\end{equation}
Then, for the relaxation under the inverse tempered $\alpha$-stable process the function
(\ref{eq18}) takes the form
\begin{equation}
\chi(\omega)=\frac{1}{1-\sigma^\alpha+(i\omega/\omega_p+\sigma)^\alpha}\,,\label{eq19}
\end{equation}
where $0\leq\sigma\leq 1$ is a constant, and $\omega_p$ is the characteristic frequency
of the relaxing system.

According to Eq. (\ref{eq19}), for $\sigma=0$ the relaxation function describes the
Cole-Cole law. If $\alpha=1$, the function becomes exponential. In the case of $\sigma=1$
it has the Cole-Davidson form. The relaxation directed by the inverse tempered
$\alpha$-stable process takes an intermediate place between the superslow relaxation and
the exponential one (see Fig.~\ref{fig:relax}). Such a type of evolution is observed in
relaxation experiments (see, for example, \cite{J2}).

{\it Conclusions.}--- In summary we have developed a novel approach to anomalous
diffusion and nonexponential relaxation from tempered $\alpha$-stable processes. The
model is broader than the purely subdiffusive case. It is very important that they both
can be considered on the unique base following the theory of subordinated random
processes. We have derived a tempered form of the FPE and the relaxation function, as
well as we have calculated the mean square displacement describing the processes. In
Fig.~\ref{fig:prop}, as an example, the propagator $p(x,t)$ for the tempered diffusion
with $\alpha$ = 2/3 and $\delta$ = 0.5, is drawn. The cusp shape of the p.d.f. disappears
when time increases. Thus our model occupies an intermediate place between subdiffusion
and normal diffusion. We expect that our results will yield insights into the coexistence
of subdiffusion and normal diffusion in nature.

AS is grateful to the Institute of Physics, Wroc{\l}aw University  of Technology 
for kind hospitality during his visit. 

\bibliographystyle{apsrev}
\bibliography{stanislavsky}

\end{document}